\documentclass[10pt,conference]{IEEEtran}
\IEEEoverridecommandlockouts

\usepackage{cite}
\usepackage{amsmath,amssymb,amsfonts}
\usepackage{algorithmic}
\usepackage{graphicx}
\usepackage{textcomp}
\usepackage{xcolor}
\usepackage{url}
\usepackage{array}
\usepackage{multirow}
\usepackage{makecell}
\usepackage{tabularx}
\usepackage{booktabs}

\def\BibTeX{{\rm B\kern-.05em{\sc i\kern-.025em b}\kern-.08em
T\kern-.1667em\lower.7ex\hbox{E}\kern-.125emX}}

\begin{document}

\title{Compact Convolutional Segmentation for Visual Landmark Extraction in GNSS-Denied UAV Navigation}

\author{
Sude Ertan\IEEEauthorrefmark{1} , Osman Tokluoglu\IEEEauthorrefmark{1}, Mustafa Ozturk\IEEEauthorrefmark{2}\\
\IEEEauthorrefmark{1}Department of Electrical and Electronics Engineering, 
Ankara Yildirim Beyazit University, Ankara, Turkiye\\
\IEEEauthorrefmark{2}TAYF Research and Development Engineering Consulting Inc., Ankara, Turkiye\\
suddertan@gmail.com, otokluoglu@aybu.edu.tr, mustafa@tayfargem.tr
}
\IEEEpubid{
\makebox[\columnwidth]{%
\textbf{979-8-3195-4709-5/26/\$31.00 \copyright 2026 IEEE}%
\hfill}
\hspace{\columnsep}
\makebox[\columnwidth]{}
}
\maketitle

\begin{abstract}
Reliable localization of unmanned aerial vehicles
(UAVs) becomes challenging when Global Navigation Satellite
System (GNSS) signals are degraded, blocked, or intentionally
jammed. In such GNSS-denied conditions, visual information
obtained from onboard cameras can provide complementary
cues for navigation by identifying spatially stable and distinc
tive landmarks. This study proposes a compact convolutional
segmentation framework for extracting candidate visual land
marks from aerial imagery. The proposed model combines
fully convolutional processing with dilation-based spatial con
text extraction and residual feature transfer. Since a dedicated
UAV landmark dataset is not available in this study, an aerial
building segmentation dataset is adapted as an initial evaluation
environment. Experimental results indicate that the proposed
architecture provides a feasible front-end for candidate landmark
extraction, while further improvements are required through
extended training, UAV-specific datasets, and integration with
localization or matching algorithms..
\end{abstract}

\begin{IEEEkeywords}
FCN, GNSS, landmark extraction, ResNet, UAV navigation, U-Net.
\end{IEEEkeywords}

\section{Introduction}
\label{sec:introduction}

Unmanned aerial vehicles (UAVs) increasingly rely on autonomous localization capabilities during monitoring, inspection, and surveillance missions. However, this capability can be severely affected when Global Navigation Satellite System (GNSS) signals are unavailable, degraded, or intentionally jammed. GNSS-denied UAV navigation has therefore been investigated through inertial navigation, LiDAR-based localization, visual navigation, image matching, and landmark-based path planning~\cite{perez2018architecture,lu2018survey,balamurugan2016survey,hemann2016long,singh2016landmarks,Tokluoglu2019IMU,braga2016image,goforth2019gps}. Although inertial and LiDAR-based approaches can provide useful localization support, they may suffer from drift accumulation, increased hardware cost, or additional system complexity. In this context, camera-based visual landmark extraction provides a practical alternative, particularly for UAVs that already carry onboard cameras for monitoring, inspection, and surveillance missions.

Landmark-based navigation is meaningful for GNSS-denied UAV operation because stable visual structures can act as reference points when absolute positioning information is unavailable. Singh and Sujit~\cite{singh2016landmarks} investigated landmark-based path planning for UAVs in GPS-denied areas, while image matching-based studies~\cite{braga2016image,goforth2019gps} showed that aerial or satellite imagery can support UAV localization. These studies motivate the extraction of visually distinguishable and spatially stable regions from aerial imagery.

Recent studies further indicate that visual localization and geo-localization remain active research topics for UAVs operating under degraded or unavailable GNSS conditions. Deep learning-based visual localization methods have been reviewed as promising alternatives to conventional template matching and handcrafted feature matching approaches for GPS-denied UAV navigation~\cite{aljarrah2024exploring}. Similarly, recent UAV geo-localization surveys emphasize the role of matching UAV imagery with geo-referenced satellite or aerial maps, particularly when direct GNSS measurements are unreliable or unavailable~\cite{avola2024survey}. Beyond survey studies, recent works have addressed practical challenges such as viewpoint discrepancy, oblique camera geometry, aerial-to-satellite appearance differences, and multi-source localization fusion. Chen and Jiang~\cite{chen2024oblique} investigated oblique-robust UAV-to-satellite visual localization, while Zhou et al.~\cite{zhou2025satloc} introduced the SatLoc dataset and a hierarchical adaptive fusion framework for GNSS-denied UAV localization. In addition, Ostrovskyi et al.~\cite{ostrovskyi2025landmarks} studied distinctive landmark discovery from earth imagery using deep feature outliers for robust UAV geo-localization. These recent studies show that reliable UAV localization depends not only on matching or fusion algorithms, but also on the availability of stable, distinctive, and repeatable visual structures. Therefore, extracting stable landmark regions from aerial imagery remains a relevant front-end problem for future GNSS-denied UAV navigation systems.

Landmark detection has also been studied in several related domains, including robotic navigation~\cite{song2012natural,liu2010natural}, indoor and small-scale outdoor navigation~\cite{perez2018architecture,scaramuzza2014vision}, acoustic landmark extraction~\cite{wijk2000localization,ko1996method}, social media image analysis, facial landmark detection, underwater visual SLAM, and satellite image navigation~\cite{samany2019automatic,bodini2019review,anderson2012robot,aulinas2011feature,kim2005landmark}. Geometric landmark extraction using Voronoi-based segmentation has also been proposed~\cite{zhang2018landmark}. Although these studies demonstrate the broader relevance of landmark extraction, UAV-based landmark extraction from aerial flight images introduces specific difficulties such as scale variation, shadows, background clutter, and viewpoint-dependent changes.

Stable man-made structures may serve as candidate visual landmarks for UAV navigation. Their extraction from aerial imagery is challenging because of scale, illumination, and background variations, motivating a segmentation-based approach.

Building segmentation and multi-scale aerial image segmentation studies show that convolutional architectures can learn meaningful structural patterns from overhead imagery~\cite{maggiori2016fully,Yi2019BuildingSO,ji2019scale,li2019aerial}. At the architectural level, fully convolutional networks enable dense pixel-level prediction without fully connected layers~\cite{long2015fully}, while U-Net-like encoder-decoder structures preserve spatial details through feature transfer between downsampling and upsampling stages~\cite{ronneberger2015u}. Residual learning further supports feature propagation and mitigates degradation in deeper networks~\cite{he2016deep}. However, the use of such extracted regions as visual landmarks for GNSS-denied UAV navigation still requires further investigation.

In this study, a compact fully convolutional segmentation architecture is proposed for landmark extraction from aerial imagery. The proposed model is designed as a visual front-end rather than a complete UAV navigation system, and the extracted landmark-like regions are intended to support later matching, localization, map association, or route-planning stages. Within this scope, dilation-based spatial context extraction and residual feature transfer are combined in a lightweight architecture, and their effects are evaluated through an ablation-style comparison with plain, dilation-based, residual, and combined variants under the same training configuration.

The remainder of this paper is organized as follows. Section~\ref{blm:yontem} presents the methodology, including the proposed approach, dataset, and network architecture. Section~\ref{blm:results} discusses the experimental results and comparative evaluation. Finally, Section~\ref{blm:conclusion} concludes the paper and outlines future work.

\section{Method}
\label{blm:yontem}

This section presents the methodology followed in the study. First, the general approach is described by positioning landmark extraction as a segmentation-oriented front-end for UAV navigation. Then, the dataset and training configuration are introduced. Finally, the proposed architecture is explained.

\subsection{Approach}
\label{blm:a_aproach}

This study investigates the segmentation of static man-made structures as candidate visual landmarks in aerial imagery. The method is intended as a preliminary front-end for later matching or localization stages, rather than a complete UAV navigation solution.

The proposed approach follows the principles of fully convolutional segmentation. FCN-based models generate dense pixel-level predictions while preserving the spatial organization of the input image~\cite{long2015fully}. U-Net-like encoder-decoder structures support the recovery of spatial details through feature transfer between downsampling and upsampling stages~\cite{ronneberger2015u}, while residual connections improve feature propagation and reduce degradation in deeper models~\cite{he2016deep}. Based on these motivations, the proposed architecture combines FCN-based dense prediction, dilation-based spatial context extraction, and residual feature transfer within a compact structure.

Dilation is used to enlarge the effective receptive field without substantially increasing the number of parameters, which is useful for capturing landmark regions at different apparent scales. Residual connections are incorporated to preserve earlier feature information and support spatial detail recovery after downsampling. Batch normalization is applied after convolution operations to stabilize training, and dropout is used in selected activation layers as a regularization mechanism. Unlike conventional architectures that commonly use channel sizes based on powers of two, this study explores a compact channel configuration inspired by selected Fibonacci and prime-number-related values as a practical design choice rather than a theoretical optimum.

\subsection{Dataset and Training}
\label{blm:a_dataset}

The dataset introduced by Mnih~\cite{MnihThesis} was used for training and evaluation. This dataset contains aerial images and corresponding segmentation masks for road and building regions from Boston and Massachusetts. Since a dedicated UAV landmark extraction dataset was not available in this study, the building segmentation data were adapted as an initial evaluation environment. This choice is motivated by the structural similarity between building segmentation and landmark extraction, since buildings and similar static objects can serve as visually distinguishable reference regions for UAV navigation.

\begin{figure}[!t]
\centering
\includegraphics[width=\linewidth]{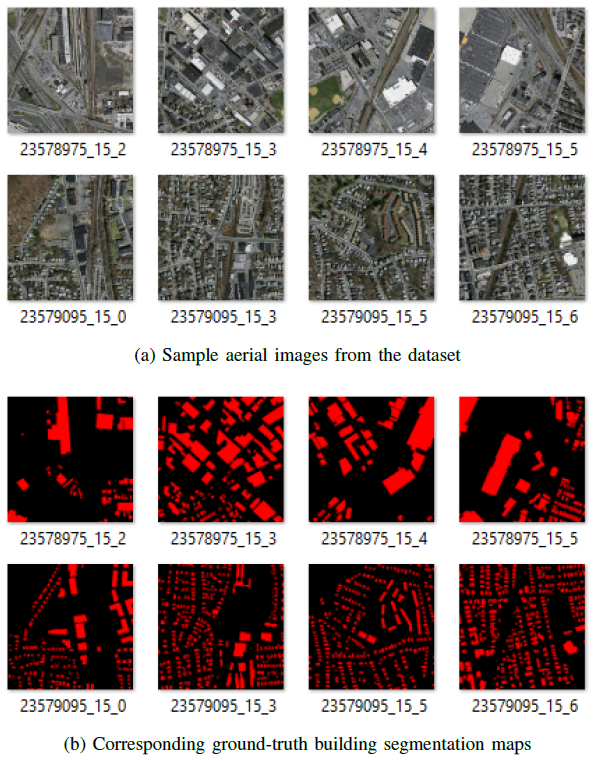}
\caption{The dataset used for network training was adapted from~\cite{MnihThesis}. (a) Sample input aerial images; (b) corresponding ground-truth building segmentation maps.}
\label{fig:Fig_1}
\end{figure}

The original images have a resolution of $1500 \times 1500$. They were divided into nine non-overlapping sub-images of size $500 \times 500$ to make them compatible with the proposed training pipeline. Images with unsuitable or irrelevant content were removed after this partitioning stage. Before being fed into the network, the selected $500 \times 500$ sub-images were resized to $192 \times 192$ pixels to match the input size of the proposed architecture. The image samples were resized using interpolation, while the corresponding segmentation masks were resized with label-preserving nearest-neighbor interpolation. The final dataset consisted of 600 training images, 63 validation images, and 26 test images. Sample image-mask pairs are shown in Fig.~\ref{fig:Fig_1}. Fig.~\ref{fig:Fig_1}(a) shows sample aerial input images, while Fig.~\ref{fig:Fig_1}(b) presents the corresponding ground-truth building segmentation masks.

During training, the Adam optimizer was used with a learning rate of 0.005. The batch size was set to 200, and the number of epochs was set to 10. The same training configuration was applied to all compared model variants in order to provide a controlled architectural comparison.

\subsection{Proposed Method}
\label{blm:a_proposed}

\begin{figure*}[!t]
\centering
\includegraphics[width=0.7\textwidth]{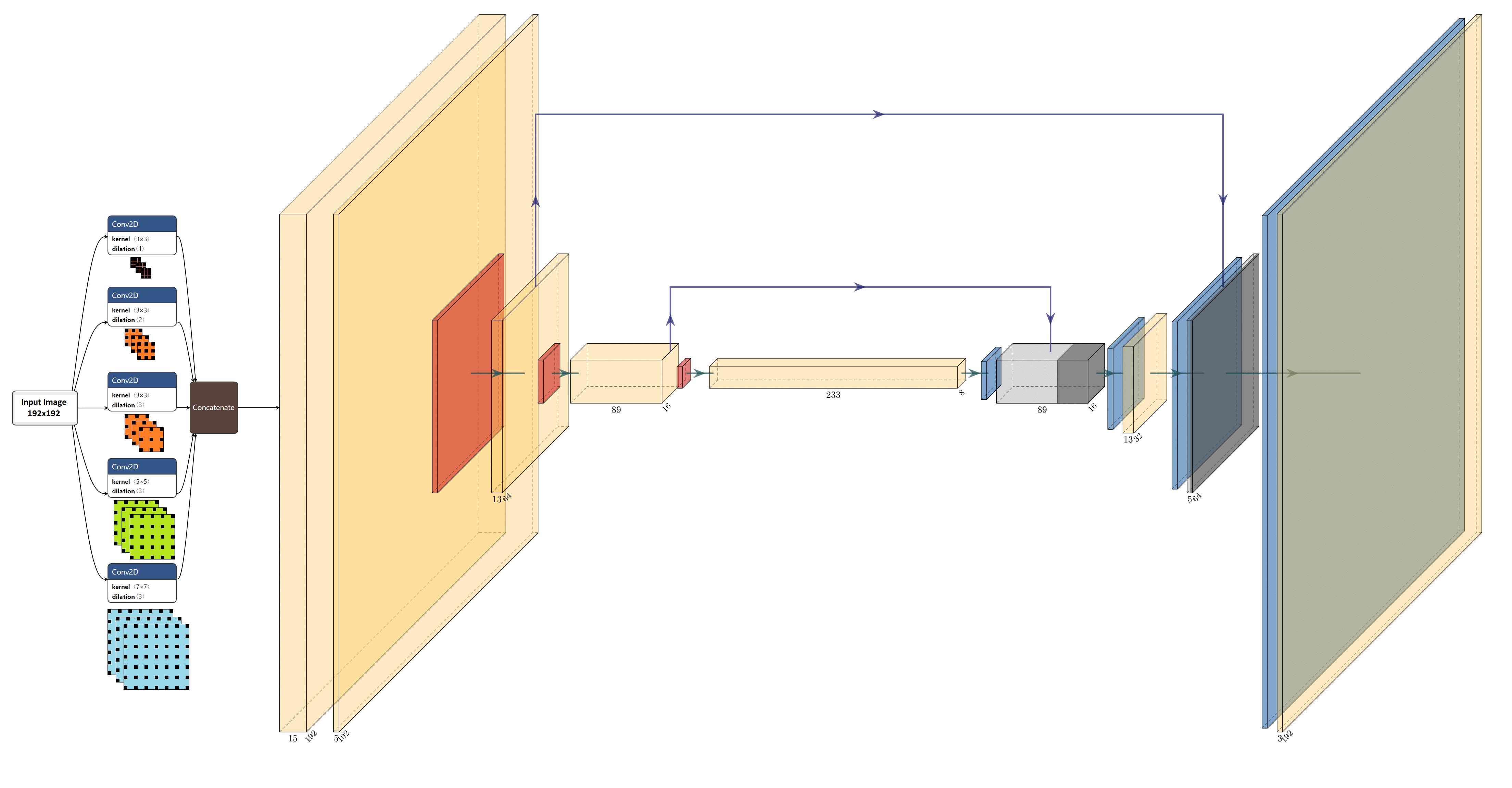}
\caption{Proposed deep convolutional neural network architecture. A $192 \times 192$ input image is processed through dilated convolutional branches to capture multi-scale representations. The extracted features are refined through a compact U-shaped structure with skip and residual connections to support spatial information transfer and feature propagation.}
\label{fig:Fig_2}
\end{figure*}

The proposed network is a compact convolutional segmentation model composed of nine convolutional layers. The overall structure is illustrated in Fig.~\ref{fig:Fig_2}. ReLU is used as the activation function in the convolutional stages. Batch normalization is applied after convolution operations up to the downsampling stage, and the normalized feature maps are subsequently passed through nonlinear activation layers.

The architecture follows a U-shaped segmentation structure. Downsampling is performed through max-pooling until the bottleneck stage is reached. The network then reconstructs spatial information through an expansion path. After the initial expansion stage, selected channel values are used to keep the model compact. In the implemented configuration, the first convolutional stage contains 5 channels, while subsequent stages use 23, 89, and 233 channels, respectively. These values are selected as a compact non-standard channel configuration rather than as a claim of mathematical optimality.

The architecture can be divided into two main parts. The first part uses dilation-based convolutional processing to enlarge the receptive field and capture spatial structures at different scales. The second part follows a modified U-Net-like structure with skip and residual connections to support feature propagation and preserve spatial detail. The combination of these mechanisms is intended to improve the extraction of building-like landmark regions while keeping the network relatively lightweight.

The model is designed with UAV-oriented computational constraints in mind. Since onboard UAV platforms may have limited processing capability, compactness is an important consideration. However, the present study does not provide a full embedded deployment analysis. Therefore, the proposed model should be interpreted as a compact candidate architecture whose suitability for real-time onboard operation requires further validation through inference-time, memory, and energy measurements on embedded hardware.

\begin{table}[t]
\renewcommand{\arraystretch}{1.1}
\caption{Summary of the proposed network stages.}
\label{tabArch}
\centering
\resizebox{\columnwidth}{!}{%
\begin{tabular}{l c c}
\hline
Stage & Operation & Purpose \\
\hline
Input & $192 \times 192$ RGB image & Aerial patch representation \\
Context block & Dilated convolution branches & Multi-scale spatial context \\
Encoder & Convolution + pooling & Compact feature extraction \\
Bottleneck & Convolutional layers & Low-dimensional representation \\
Decoder & Upsampling + skip/residual links & Spatial detail recovery \\
Output & $1 \times 1$ convolution & Segmentation map \\
\hline
\end{tabular}%
}
\end{table}

\section{Results and Discussion}
\label{blm:results}

Four model variants were trained and evaluated using the constructed dataset, and the results are presented in Table~\ref{tabCompTest}. IoU, precision, and recall were computed at the pixel level for the target building/landmark class after thresholding the predicted segmentation maps. Model 0 is the plain convolutional baseline with batch normalization, pooling, and dropout operations. Model 1 adds dilation-based convolutional processing, Model 2 introduces residual connections, and the proposed method combines both mechanisms within a compact FCN-based segmentation structure.

\begin{table}[t]
\renewcommand{\arraystretch}{1.2}
\caption{Comparison of models based on training and test results.}
\label{tabCompTest}
\centering
\resizebox{\columnwidth}{!}{%
\begin{tabular}{|c|c|c|c|c|c|c|}
\hline
Method & Train/Test & Loss & Accuracy & IoU & Precision & Recall \\
\hline
Model 0 (Plain) & Train & 0.1233 & 0.4948 & 0.1150 & 0.1045 & 0.0394 \\
                & Test  & 0.1599 & 0.3824 & 0.1108 & 0.1099 & 0.0289 \\
\hline
Model 1 (Dilation) & Train & 0.1245 & 0.5512 & 0.1158 & 0.1065 & 0.0380 \\
                   & Test  & 0.1833 & 0.4877 & 0.1304 & 0.1518 & 0.0566 \\
\hline
Model 2 (Residual) & Train & 0.1176 & 0.5504 & 0.1304 & 0.1473 & 0.0823 \\
                   & Test  & 0.1734 & 0.4877 & 0.1381 & 0.2202 & 0.0862 \\
\hline
Proposed Method & Train & 0.1146 & 0.4946 & 0.1347 & 0.1568 & 0.0909 \\
                & Test  & 0.1468 & 0.3824 & 0.1324 & 0.1721 & 0.0803 \\
\hline
\end{tabular}%
}
\end{table}

The comparison provides an ablation-style evaluation of the architectural components under the same training conditions. Model 1 is used to observe the effect of dilation-based spatial context extraction, while Model 2 is used to evaluate the contribution of residual feature transfer. Therefore, Table~\ref{tabCompTest} should be interpreted as an internal architectural comparison rather than a fully optimized benchmark against state-of-the-art segmentation networks.

According to Table~\ref{tabCompTest}, the proposed method achieves the lowest loss on both the training and test sets. On the training set, it also provides the highest IoU, precision, and recall values among the evaluated models. On the test set, however, Model 2 obtains higher IoU, precision, and recall values, while the proposed method maintains the lowest test loss. This indicates that residual connections are particularly beneficial for preserving spatial information and improving segmentation overlap. At the same time, the lower test loss of the proposed method suggests that the combination of dilation and residual processing provides a stable compact architecture under the selected training configuration.

The results also show that model performance should not be evaluated using a single metric. In segmentation tasks, IoU, precision, and recall provide complementary information about region overlap, false detections, and missed landmark regions. The relatively low recall values indicate that some target regions are not sufficiently detected, which may be related to the limited number of epochs, the small test set, and the lack of UAV-specific training data. Therefore, the current results are best interpreted as a feasibility-oriented architectural evaluation rather than a finalized landmark extraction system.

\begin{figure}[!t]
\centering
\includegraphics[width=\linewidth]{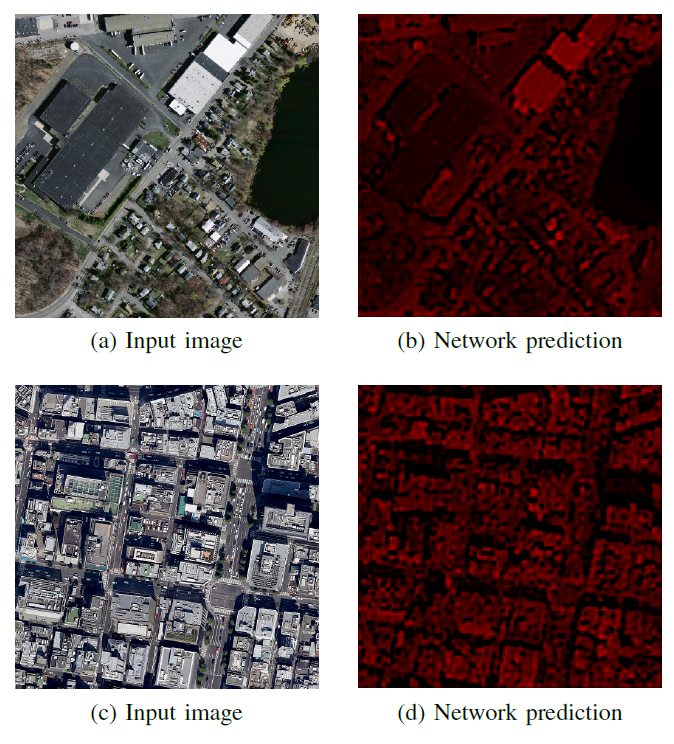}
\caption{The input images shown in (a) and (c) were used to evaluate the proposed method. The corresponding output predictions obtained from the network are presented in (b) and (d).}
\label{fig:Fig_3}
\end{figure}

Fig.~\ref{fig:Fig_3} presents qualitative predictions obtained from aerial images. The model mainly responds to building-like structural patterns, while non-landmark regions such as water, roads, and open land are less explicitly emphasized. This behavior is consistent with the training objective based on building segmentation masks. However, the visual results also indicate that additional optimization is required to improve boundary sharpness, region completeness, and detection consistency. Longer training, stronger data augmentation, class-imbalance-aware loss functions, and UAV-specific annotated images may improve the robustness of the extracted landmark regions.

The effect of input size was also examined. When the input resolution was increased, no substantial improvement was observed under the current training setting. Therefore, $192 \times 192$ was retained as the default input size. Convolution operations were performed with padding in order to preserve spatial dimensions. Although this helps maintain feature-map size, boundary effects may still influence segmentation quality near image edges. Alternative padding strategies, deeper convolutional blocks, or multi-scale supervision may be considered in future work.

Overall, the findings suggest that the proposed architecture can serve as a compact front-end for landmark extraction. Since the dataset is not UAV-specific and the study does not yet include localization, matching, or onboard inference experiments, future evaluation should consider complete UAV localization pipelines together with computational metrics such as parameter count, inference time, memory usage, and frame rate.
\section{Conclusion}
\label{blm:conclusion}
This study presented a compact FCN-based segmentation framework for extracting candidate visual landmarks from downward-facing aerial imagery under GNSS-denied UAV navigation scenarios. Since no dedicated UAV landmark dataset was available, an existing aerial building segmentation dataset was adapted as an initial evaluation environment. The results indicate that segmentation-based feature learning can provide a feasible visual front-end for landmark extraction; however, further optimization is needed to improve segmentation overlap, recall, and generalization. Future work will address extended training, UAV-specific annotated datasets, architectural refinements, and integration with matching, localization, or map-association algorithms. Although the lightweight design suggests potential for embedded real-time deployment, this should be validated through onboard inference-time, memory, frame-rate, and energy measurements. FPGA-based implementation with quantization and pruning may further support low-latency, energy-efficient onboard processing.

\end{document}